\def\ugr{\lower4pt \hbox{$\buildrel > \over \sim$}}
\def\ukl{\lower4pt \hbox{$\buildrel < \over \sim$}}

\documentstyle[psfig, epsf, rotate]{l-aa}
\begin{document}

\thesaurus{}
\title{The radio spectra of galactic nuclei}
\author{M.  B\"ottcher \inst{1, 2} \and H.-P. Reuter \inst{2}
\and H. Lesch \inst{3}}
\institute{Rice University, Space Physics and Astronomy Department,
6100 S. Main Street, Houston, TX 77005, USA
\and
Max-Planck-Institut f\"ur Radioastronomie, Postfach 20 24,
D -- 53 010  Bonn, Germany
\and
Institut f\"ur Astronomie und Astrophysik der Universit\"at M\"unchen,
Scheinerstra\ss e 1, D - 81 679 M\"unchen, Germany}
\date{Received 22 April 1997; Accepted 31 July 1997}
\offprints{M. B\"ottcher}

\maketitle

\begin{abstract}

We present a model for the inverted radio spectra of
active active galactic nuclei as well as the central 
regions of normal galaxies.
The model is based on the unified scenario for active 
galaxies, stating that the central engines of active 
galaxies consists of a supermassive black hole surrounded 
by an accretion disk and a radio jet. The nuclei of normal 
(i. e. less active) galaxies are supposed to be scaled-down 
versions of the same phenomenon. We show that the radio 
emission of a jet component, becoming optically thin to the 
radio emission of a monoenergetic pair plasma at decreasing
frequencies as it moves outward and expands, is well suited 
to explain the observational results. We present a model 
calculation for the special case of the nucleus of M~81.

\end{abstract}

\section{Introduction}

Radio observations of many active galactic nuclei (AGNs) as well 
as the nuclei of normal galaxies in the MHz through GHz frequency 
range reveal inverted radio spectra with spectral indices
$0 \le \alpha \le 0.4$ ($S_{\nu} \propto \nu^{-\alpha}$)
(e. g., K\"uhr et al. 1981, Stickel et al. 1991). In 
the case of NGC~3031 (M~81), the observed radio 
spectrum is consistent with optically thin
synchrotron emission of a monoenergetic distribution of
relativistic electrons, i. e. a spectrum of 
the shape $S_{\nu} \sim \nu^{1/3} \, e^{-\nu / \nu_c}$, 
as was recently suggested by Reuter \& Lesch (1996).

Previously, Bartel et al. (1982) proposed a model 
in which the inverted $\nu^{1/3}$ spectrum is assumed 
to be the emission of an inhomogeneous, optically thick
source with the magnetic field and electron energy
distribution and density varying radially. As was argued 
in Reuter \& Lesch (1996), this model requires a rather
unlikely magnetic field value ($B \approx 200$~G) and
an improbable fine tuning of the radial dependencies
to reproduce the $\nu^{1/3}$ spectrum.

In a more recent paper, Falcke (1996) discussed a
model involving optically thin emission of plasma of
monoenergetic particles in a continuously filled radio 
jet. He emphasized the effect of the pressure gradient 
along the jet leading to an acceleration of the bulk
motion, which was claimed to be responsible for the
inverted shape of the radio spectrum. For the special
case of NGC~3031, Doppler enhancement had to
be invoked to account for the observed intensity which,
however, appears to be unlikely given the large viewing 
angle of $i \approx 60^{\circ}$ relative to the plane of 
the entire galaxy.

We favour a similar idea, suggesting that the 
inverted radio spectra arise from the synchrotron 
emission of expanding radio jet components.
In contrast to the papers by Falcke (1996) and 
by Reuter \& Lesch (1996), we argue that 
in the case of NGC~3031 it is more likely 
that the synchrotron emission is not optically
thin over the whole range of frequencies. We 
assume that the jet is not filled continuously
with relativistic plasma but that single, separated
relativistic blobs move through the jet cone.
This is indicated by the correlation of $\gamma$-ray 
flares of blazar type AGNs with the ejection of new 
radio components from their core regions (e. g., Britzen 
et al. 1996). Furthermore, we assume that that 
hydromagnetic turbulences counteract radiative 
and adiabatic energy losses of the particles 
in the jet components. 

The unified scenario for active galactic nuclei 
(e. g. Urry \& Padovany 1995) states that the 
central engines of active galaxies
consist of a supermassive black hole ($M \sim 10^6$
-- $10^9 \, M_{\odot}$), surrounded by an accretion
disk and accelerating a jet of relativistic particles
perpendicular to the disk plane. Models based on
this idea are well suited to explain, e. g.
the broadband radiation and variability of Seyfert 
galaxies, quasars and BL~Lac~objects (e. g. Svensson 1997, 
Dermer, Sturner \& Schlickeiser 1997, Bloom \& Marscher 1996). 
However, jet models for AGNs usually fail to reproduce the 
radio emission of these objects in the cm to mm range since 
the emitting regions responsible 
for the optical through $\gamma$-ray emission, are
optically thick to synchrotron emission in the
radio frequency range.

We assume that in a single component of a relativistic
jet ejected by a central engine, after the initial phase 
in which high-energy emission can be produced, a 
narrow particle distribution with relativistic 
mean energy $\langle \gamma \rangle \gg 1$ is 
built up (B\"ottcher et al. 1997). As the jet moves 
outwards, it expands and the system becomes optically 
thin to synchrotron emission at lower frequencies. We
demonstrate that the sum of the synchrotron spectra 
of a few such components is in agreement with the 
observed inverted radio spectra and specify this 
idea to the case of NGC~3031.

\section{The model}

Theoretical investigations on the broadband spectra of
AGNs based on relativistic jet models indicate that jets 
in AGNs most probably contain a relativistic pair plasma
where the particles have Lorentz factors above some 
cutoff $\gamma_{min}$ (e. g., Sikora 1997, Dermer et 
al. 1997, Bloom \& Marscher 1996). This
low-energy cutoff corresponds to the pair-production 
threshold if the pairs filling the jet components 
are of secondary origin. The combined action of 
synchrotron and inverse-Compton losses will then 
lead to the establishment of a very narrow particle 
distribution (B\"ottcher et al. 1997, Achatz et 
al. 1990) centered at relativistic energies. Such 
a distribution is called quasi-monoenergetic.
After nearly complete dilution of the source 
photon fields for Compton scattering (synchrotron or 
external photons) the further evolution of the particles 
inside such a jet component will be governed by the 
structure of the global magnetic field as well as
synchrotron losses and reacceleration by hydromagnetic
turbulences. As long as the jet does not widen
significantly, these relativistic plasma components
are optically thick to synchrotron emission up to
frequencies near the critical frequency $\nu_c \, = 
\, {3 \over 2} \, \nu_e \, \gamma^2 \, = \, 
4.2 \cdot 10^6 \, {\rm Hz} \> B \, \gamma^2$,
where $\nu_e$ is the nonrelativistic gyrofrequency and
$\gamma$ is the electron Lorentz factor. As the plasma 
component moves further out, the magnetic field is supposed 
to decrease. At some point, it will become too weak to
compensate the dynamical pressure of the relativistic
plasma, and the jet starts to widen up. If no particles
are created along the jet, this expansion leads to a 
reduction of the optical depth to synchrotron self 
absorption which results in the source becoming optically
thin at lower frequencies. At the same time the magnetic 
field decreases, implying a decrease of the critical
synchrotron frequency of the relativistic particles.

Our basic idea is that the nuclei of normal 
galaxies contain scaled-down versions 
of the same phenomenon as AGNs, i. e. a (not necessarily 
less massive) black hole, surrounded by a less luminous 
accretion disk and ejecting a less powerful relativistic jet.
The analogy between the core regions of powerful radio galaxies 
and the radio cores of normal elliptical and S0 galaxies has
been studied in detail by Slee et al. (1994), demonstrating
that these objects have similar radio properties.

In order to test the applicability of this model, we consider 
a simplified case which illustrates the basic concept: We start
our calculation at the point where by means of the scenario
outlined above a quasi-monoenergetic pair distribution inside 
a relativistic jet component has been established. Initially,
this component, containing electrons and positrons of energy 
$\gamma_0 m_e c^2$, is located at a distance $z_0$ from the 
core of the galaxy. It starts to expand freely at the point
where the magnetic field becomes too weak to collimate the jet 
structure. We assume that at this point equipartition holds.
The component is supposed to move relativistically along a 
conical jet ($z$ direction). The magnetic field declines as 
$B(z) = B_0 \, (z / z_0)^{-1}$ (Blandford \& K\"onigl 1979). In 
the phase of free expansion, the radius $R$ of the component increases 
linearly, i. e. $R(z) = R_0 \, (z / z_0)$. We neglect the escape 
of particles and assume that no particles are created or annihilated. 
Thus, since we assume that the component expands in all three 
dimensions, the density $n$ varies as $n(z) = n_0 \, (z / z_0)^{-3}$. 
The combined effect of synchrotron losses, adiabatic losses and
reacceleration by plasma wave turbulences will, in general, lead to 
some change in the electron energy. We assume that this cooling
(or heating, respectively) is described by $\langle\gamma\rangle (z) 
= \gamma_0 (z / z_0)^{-s}$ where $s$ is considered as a free parameter. 
From the value of $s$ which yields the best agreement with the data, one 
can determine the importance of stochastic acceleration as compared 
to radiative and adiabatic losses.

The synchrotron spectrum of a narrow relativistic particle 
distribution centered at $\langle\gamma\rangle$ does not differ 
much from the spectrum of a monoenergetic distribution at energy $\langle\gamma\rangle$ (e. g., Beckert et al. 1997). Therefore 
we may approximate the emanating synchrotron radiation by the
emission of monoenergetic particles. For this purpose, we
use the solution given by Crusius \& Schlickeiser (1988) for the
case that plasma effects (Razin-Tsytovich effect) on the 
synchrotron emission are negligible. The emissivity is
then given by

$$
\epsilon_{\nu} = {n \, \alpha \, h \over 4 \sqrt 3 \, \gamma^2} \, \nu
\> CS \left( {\nu \over \nu_c} \right)
$$
\begin{equation}
\approx \> 2.09 \cdot 10^{-22} \, {{\rm erg} \over {\rm s \, Hz \, 
sr \, cm}^3} \; B^{2/3} \> n \> \gamma^{- 2/3} \> \nu_9^{1/3} \, 
e^{-{\nu \over \nu_c}}
\end{equation}
where $B$ is in G, $n$ in cm$^{-3}$ and $\nu_9 = \nu / (1 {\rm GHz})$.
$\alpha$ is the hyperfine structure constant, $h$ is the Planck
constant, and

\begin{equation}
CS (x) = W_{0, {4 \over 3}} (x) W_{0, {1 \over 3}} (x) -
W_{{1 \over 2}, {5 \over 6}} (x) W_{-{1 \over 2}, {5 \over 6}} (x)
\end{equation}
where $W_{\mu, \nu} (x)$ denote the Whittaker functions.
The absorption coefficient is

$$ 
\kappa_{\nu} = {n \, \alpha \, h \over 4 \sqrt 3 \, m_e \, \gamma^3 \, \nu}
\> W_{{1 \over 2}, {5 \over 6}} \left( {\nu \over \nu_c} \right)
\, W_{-{1 \over 2}, {5 \over 6}} \left( {\nu \over \nu_c} \right)
$$
\begin{equation}
\approx \> 1.53 \cdot 10^{-13} \, {\rm cm}^{-1} \> B^{2/3} \> n \> 
\gamma^{-5/3} \> \nu_9^{-5/3} \, e^{-{\nu \over \nu_c}}
\end{equation}
Since the component is assumed to move relativistically
relative to the observer, Doppler boosting must be taken into
account. Let $\Gamma$ be the Lorentz factor of the bulk
motion, $\beta_{\Gamma} = \sqrt{1 - \Gamma^{-2}}$, and $\mu$
the angle cosine between the jet axis and the line of sight.
Then, the flux received from one plasma component located at a 
distance $z$ from the central engine, is

\begin{equation}
S_{\nu} (z) \approx D^3 \, {R(z)^2 \over d^2} {\epsilon_{\nu_i} (z) 
\over \kappa_{\nu_i} (z)} \> \left( 1 - e^{- \tau_{\nu_i} (z)} \right)
\end{equation}
where $D = \bigl( \Gamma \, [1 - \beta_{\Gamma} \mu] \bigr)^{-1}$ and
$\nu_i = \nu / D$, $d$ is the distance to the source and $\tau_{\nu_i} 
(z) \approx \kappa_{\nu_i} (z) \, R$.

\subsection{Application to M81}

In order to test our idea in the special case of NGC~3031, we use 
the same data as given in Reuter \& Lesch (1996). VLBI measurements
revealed a $\nu^{-0.8}$ dependence of the extent of the source
(Bietenholz et al. 1995), being $\approx 1000$~AU at 8.3~GHz. The 
radio emission of NGC~3031 exhibits strong variability on timescales 
of several weeks to months (de Bruyn et al. 1976).

NGC~3031 is seen under an inclination angle of $i \approx 60^{\circ}$.
We expect that the orientation of the jet structure does not deviate much
from the angular momentum vector of the entire galaxy. Therefore, an 
inclination angle $\theta \ugr 50^{\circ}$ and a moderate Doppler factor 
of $0.1 \ukl D \ukl 1$ seem plausible.

With the parameters found by Reuter \& Lesch (1996) for a non-moving 
source in a steady state in the nucleus of NGC~3031 ($B \approx 0.4$~G, 
$\gamma \approx 400$, $R \approx 1.5 \cdot 10^{16}$~cm, 
$n \approx 20$~cm$^{-3}$), the source is optically thick up to 
$\approx 1$~GHz. We find that with the absence of Doppler enhancement
and the restriction on the size from VLBI observations it is difficult
to reproduce the radio spectrum of M81 by optically thin synchrotron 
emission down to frequencies $\nu \ukl 1$~GHz. Furthermore, a stationary
situation can not explain the frequency-dependent extent of the source.

\begin{figure}
\rotate[r]{
\epsfysize=5.5cm
\epsffile[70 20 550 500] {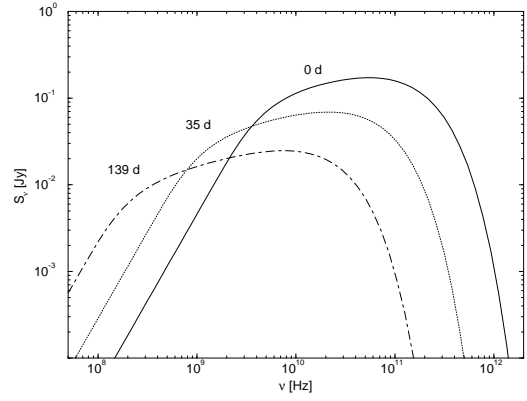}}
\caption[]{Synchrotron spectra of a single jet component 
(parameters in the text) at different times (labels). The radius of
the expanding jet component is $R = 8 \cdot 10^{15} \, {\rm cm}$, 
$2 \cdot 10^{16} \, {\rm cm}$, and $5.6 \cdot 10^{16} \, {\rm cm}$, 
respectively}
\end{figure}


Nevertheless, given the assumptions quoted above, the inverted 
spectrum is a natural consequence of the plausible fact that we
usually do not observe one single component, but the observed 
spectrum is the sum of the radio spectra from several expanding 
plasma blobs. Each blob emitts a spectrum of the form $S_{\nu}
\propto \nu^{1/3} \, e^{-\nu / \nu_c}$ above the synchrotron
self-absorption frequency, and the sum of several such components
always has a similar shape. For the model calculation illustrated
in Figs. 1 and 2, we used the following parameters: $\gamma = 250$, 
$B_0 = 1.2$~G, $R_0 = 8 \cdot 10^{15}$~cm, $n_0 = 380$~cm$^{-3}$, 
$s = 0$, and $D = 0.4$. Fig. 1 shows the temporal evolution of the 
instantaneous synchrotron emission of such a plasma component. The 
bulk of emission at a certain frequency will be received from the 
region where the component becomes optically thin to synchrotron 
emission at this frequency. For $s = 0$, this implies the following 
relation between the extent of the emitting region and the frequency:

\begin{equation}
R \propto \nu^{-{5 \over 6 - 5 s}} = \nu^{-5/6}
\end{equation}
which is in excellent agreement with the observed dependence,
$R \propto \nu^{- 0.8}$. The subsequent ejection of new 
jet components also provides a natural explanation for 
the observed variability.

\begin{figure}
\rotate[r]{
\epsfysize=5.5cm
\epsffile[70 20 550 500] {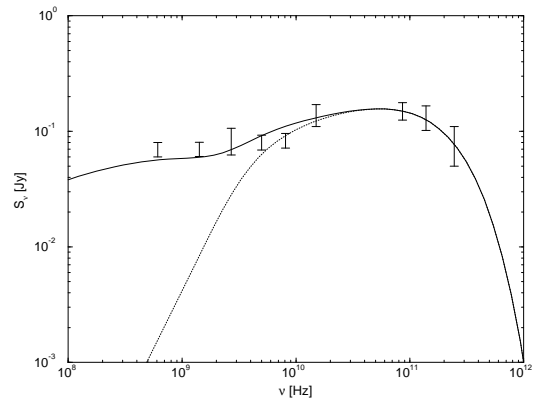}}
\caption[]{The average radio spectrum of M~81. The dotted curve 
represents the initial synchrotron spectrum (t = 0)}
\end{figure}

In Fig. 2, we compare the sum of the instantaneous synchrotron spectra
from several radio jet components, ejected with a period of a few
months, to the average radio spectrum of NGC~3031. Given the
simplicity of our assumptions, the model spectrum is in good
agreement with the observations.

\subsection{Discussion of other sources}

Radio cores of many galaxies show similar properties as observed
in M81. Recently, Muxlow et al. (1996) investigated the radio core
of the Seyfert 2 galaxy NGC~1068 between 5~GHZ and 22~GHz. They
could resolve a core component, emitting an inverted spectrum, as
well as three more extended components, probably attributable to an
expanding jet and with their spectral flux decreasing with increasing 
frequency. This is well in agreement with an evolution as described
above where the core emission represents the $S_{\nu} \propto \nu^{1/3}$
portion of an optically thin synchrotron spectrum from monoenergetic
electrons, while in the outer parts of the jet (components 1, 3 and 4
in their paper) the decreasing magnetic field has shifted the higher 
cut-off frequency of the synchrotron spectrum towards $\sim 20$~GHz.

The statistical investigations of Slee et al. (1994) demonstrate
that many E and S0 galaxies with unresolved core regions have
inverted radio spectra with spectral indices $-0.5 \ukl
\alpha \ukl 0.7$ ($S_{\nu} \propto \nu^{\alpha}$) with a mean of 
$\langle \alpha \rangle = 0.28$ which indicates the general tendency 
towards inverted spectra. Since these spectral indices are only
determined by no more than three frequencies, the influence of
high-frequency cut-offs as well as low-frequency cut-offs due
to synchrotron self-absorption or free-free absorption in
individual sources can not be fixed from these measurements.
The generally inverted radio spectra of low and medium power
with spectral indices near the value of $1/3$ may therefore 
be regarded as a hint towards synchrotron emission from 
quasi-monoenergetic electrons.

It should be noted that in the case of our galactic center,
Sgr~$A^{\ast}$, the assumption of a quasi-stationary situation
as discussed by Duschl \& Lesch (1994) and by Beckert et al. (1997) 
is more plausible. In this source, the power-law spectrum of
index $\alpha \approx 1/3$ (Zylka et al. 1995) is restricted by a 
stationary low-energy cut-off near 1~GHz, and VLBI measurements 
resolved the central source down to scales of $\sim 10^{13}$~cm 
without indicating the presence of a jet. Furthermore, the extent 
of the source is consistent with being independent of frequency. 
All these findings render it unlikely that, also in this case, we 
observe the synchrotron emission of a jet. The reason for this
is probably that there is not sufficient material in the inner 
parsec of the galaxy, usually denoted as the central cavity, to
fuel the central black hole, thus preventing the formation of
a powerful accretion disk and a jet.

\section{Summary}

Based on recent calculations about the fate of relativistic
electron distributions in the vicinity of strong sources of
radiation we can explain the physical origin of inverted
radio spectra in galactic nuclei. High energy emission
processes (inverse Compton scattering of UV-photons to 
$\gamma$-ray energies) produce a quasi-monoenergetic 
distribution function. Such quasi-monoenergetic electrons 
are injected into a plasma jet in the form of distinct 
components rather than in a continuos relativistic outflow, 
as indicated by recent observations. The time evolution 
of these components including the evolution of the optical 
depth of the radio emission of quasi-monoenergetic electrons 
along the propagating jet accounts for the observed inverted 
radio spectra.

Our model bridges the gap between the high energy behaviour of 
galactic nuclei and the radio emission. If the $\gamma$-flux of 
galactic nuclei is of purely leptonic origin via inverse Compton 
scattering on UV-photons, the observed inverted radio spectra 
are a natural consequence of the spatial and time evolution
of the relativistic electron distributions involved in both 
$\gamma$-ray and synchrotron radiation.

\end{document}